
\def\singlespace{\normalbaselines}
\def\oneandahalfspace{\baselineskip=1.15\normalbaselineskip plus 1pt
\lineskip=2pt\lineskiplimit=1pt}

\def\np{\vfill\eject}
\def\nl{\hfil\break}

\def\nofirstpagenoten{\nopagenumbers\footline={\ifnum\pageno>1\tenrm
\hss\folio\hss\fi}}
\def\nofirstpagenotwelve{\nopagenumbers\footline={\ifnum\pageno>1\twelverm
\hss\folio\hss\fi}}
\def\leaderfill{\leaders\hbox to 1em{\hss.\hss}\hfill}
\def\ft#1#2{{\textstyle{{#1}\over{#2}}}}
\def\frac#1/#2{\leavevmode\kern.1em
\raise.5ex\hbox{\the\scriptfont0 #1}\kern-.1em/\kern-.15em
\lower.25ex\hbox{\the\scriptfont0 #2}}
\def\sfrac#1/#2{\leavevmode\kern.1em
\raise.5ex\hbox{\the\scriptscriptfont0 #1}\kern-.1em/\kern-.15em
\lower.25ex\hbox{\the\scriptscriptfont0 #2}}


\parindent=20pt
\def\narrow{\advance\leftskip by 40pt \advance\rightskip by 40pt}

\def\AB{\bigskip
        \centerline{\bf ABSTRACT}\medskip\narrow}
\def\nonarrower{\advance\leftskip by -40pt\advance\rightskip by -40pt}
\def\AE{\bigskip\nonarrower}

\def\boxit#1{\vbox{\hrule\hbox{\vrule\kern3pt
        \vbox{\kern3pt#1\kern3pt}\kern3pt\vrule}\hrule}}

\def\gtorder{\mathrel{\raise.3ex\hbox{$>$}\mkern-14mu
             \lower0.6ex\hbox{$\sim$}}}
\def\ltorder{\mathrel{\raise.3ex\hbox{$<$}|mkern-14mu
             \lower0.6ex\hbox{\sim$}}}
\def\dalemb#1#2{{\vbox{\hrule height .#2pt
        \hbox{\vrule width.#2pt height#1pt \kern#1pt
                \vrule width.#2pt}
        \hrule height.#2pt}}}

\font\fourteentt=cmtt10 scaled \magstep2
\font\fourteenbf=cmbx12 scaled \magstep1
\font\fourteenrm=cmr12 scaled \magstep1
\font\fourteeni=cmmi12 scaled \magstep1
\font\fourteenss=cmss12 scaled \magstep1
\font\fourteensy=cmsy10 scaled \magstep2
\font\fourteensl=cmsl12 scaled \magstep1
\font\fourteenex=cmex10 scaled \magstep2
\font\fourteenit=cmti12 scaled \magstep1
\font\twelvett=cmtt10 scaled \magstep1 \font\twelvebf=cmbx12
\font\twelverm=cmr12 \font\twelvei=cmmi12
\font\twelvess=cmss12 \font\twelvesy=cmsy10 scaled \magstep1
\font\twelvesl=cmsl12 \font\twelveex=cmex10 scaled \magstep1
\font\twelveit=cmti12
\font\tenss=cmss10
 
 \font\ninebf=cmbx7 scaled \magstep1
\font\ninerm=cmr7 scaled \magstep1 \font\ninei=cmmi7 scaled \magstep1
\font\ninesy=cmsy7 scaled \magstep1 
\font\eightrm=cmr7 scaled 1140 
 
\font\sevenbf=cmbx7 \font\sevenrm=cmr7 \font\seveni=cmmi7
\font\sevensy=cmsy7 

\catcode`@=11
\newskip\ttglue
\newfam\ssfam

\def\fourteenpoint{\def\rm{\fam0\fourteenrm}
\textfont0=\fourteenrm \scriptfont0=\tenrm \scriptscriptfont0=\sevenrm
\textfont1=\fourteeni \scriptfont1=\teni \scriptscriptfont1=\seveni
\textfont2=\fourteensy \scriptfont2=\tensy \scriptscriptfont2=\sevensy
\textfont3=\fourteenex \scriptfont3=\fourteenex \scriptscriptfont3=\fourteenex
\def\it{\fam\itfam\fourteenit} \textfont\itfam=\fourteenit
\def\sl{\fam\slfam\fourteensl} \textfont\slfam=\fourteensl
\def\bf{\fam\bffam\fourteenbf} \textfont\bffam=\fourteenbf
\scriptfont\bffam=\tenbf \scriptscriptfont\bffam=\sevenbf
\def\tt{\fam\ttfam\fourteentt} \textfont\ttfam=\fourteentt
\def\ss{\fam\ssfam\fourteenss} \textfont\ssfam=\fourteenss
\tt \ttglue=.5em plus .25em minus .15em
\normalbaselineskip=16pt
\abovedisplayskip=16pt plus 4pt minus 12pt
\belowdisplayskip=16pt plus 4pt minus 12pt
\abovedisplayshortskip=0pt plus 4pt
\belowdisplayshortskip=9pt plus 4pt minus 6pt
\parskip=5pt plus 1.5pt
\setbox\strutbox=\hbox{\vrule height12pt depth5pt width0pt}
\let\sc=\tenrm
\let\big=\fourteenbig \normalbaselines\rm}
\def\fourteenbig#1{{\hbox{$\left#1\vbox to12pt{}\right.\n@space$}}}

\def\twelvepoint{\def\rm{\fam0\twelverm}
\textfont0=\twelverm \scriptfont0=\ninerm \scriptscriptfont0=\sevenrm
\textfont1=\twelvei \scriptfont1=\ninei \scriptscriptfont1=\seveni
\textfont2=\twelvesy \scriptfont2=\ninesy \scriptscriptfont2=\sevensy
\textfont3=\twelveex \scriptfont3=\twelveex \scriptscriptfont3=\twelveex
\def\it{\fam\itfam\twelveit} \textfont\itfam=\twelveit
\def\sl{\fam\slfam\twelvesl} \textfont\slfam=\twelvesl
\def\bf{\fam\bffam\twelvebf} \textfont\bffam=\twelvebf
\scriptfont\bffam=\ninebf \scriptscriptfont\bffam=\sevenbf
\def\tt{\fam\ttfam\twelvett} \textfont\ttfam=\twelvett
\def\ss{\fam\ssfam\twelvess} \textfont\ssfam=\twelvess
\tt \ttglue=.5em plus .25em minus .15em
\normalbaselineskip=14pt
\abovedisplayskip=14pt plus 3pt minus 10pt
\belowdisplayskip=14pt plus 3pt minus 10pt
\abovedisplayshortskip=0pt plus 3pt
\belowdisplayshortskip=8pt plus 3pt minus 5pt
\parskip=3pt plus 1.5pt
\setbox\strutbox=\hbox{\vrule height10pt depth4pt width0pt}
\let\sc=\ninerm
\let\big=\twelvebig \normalbaselines\rm}
\def\twelvebig#1{{\hbox{$\left#1\vbox to10pt{}\right.\n@space$}}}

\def\tenpoint{\def\rm{\fam0\tenrm}
\textfont0=\tenrm \scriptfont0=\sevenrm \scriptscriptfont0=\fiverm
\textfont1=\teni \scriptfont1=\seveni \scriptscriptfont1=\fivei
\textfont2=\tensy \scriptfont2=\sevensy \scriptscriptfont2=\fivesy
\textfont3=\tenex \scriptfont3=\tenex \scriptscriptfont3=\tenex
\def\it{\fam\itfam\tenit} \textfont\itfam=\tenit
\def\sl{\fam\slfam\tensl} \textfont\slfam=\tensl
\def\bf{\fam\bffam\tenbf} \textfont\bffam=\tenbf
\scriptfont\bffam=\sevenbf \scriptscriptfont\bffam=\fivebf
\def\tt{\fam\ttfam\tentt} \textfont\ttfam=\tentt
\def\ss{\fam\ssfam\tenss} \textfont\ssfam=\tenss
\tt \ttglue=.5em plus .25em minus .15em
\normalbaselineskip=12pt
\abovedisplayskip=12pt plus 3pt minus 9pt
\belowdisplayskip=12pt plus 3pt minus 9pt
\abovedisplayshortskip=0pt plus 3pt
\belowdisplayshortskip=7pt plus 3pt minus 4pt
\parskip=0.0pt plus 1.0pt
\setbox\strutbox=\hbox{\vrule height8.5pt depth3.5pt width0pt}
\let\sc=\eightrm
\let\big=\tenbig \normalbaselines\rm}
\def\tenbig#1{{\hbox{$\left#1\vbox to8.5pt{}\right.\n@space$}}}
\let\rawfootnote=\footnote \def\footnote#1#2{{\rm\parskip=0pt\rawfootnote{#1}
{#2\hfill\vrule height 0pt depth 6pt width 0pt}}}

\def\tenfoot{\tenpoint\hskip-\parindent\hskip-.1cm}
\overfullrule=0pt
\tenpoint
\def\sbullet{\raise.2em\hbox{$\scriptscriptstyle\bullet$}}
\nofirstpagenotwelve
\hsize=16.5 truecm
\baselineskip 15pt

\def\ft#1#2{{\textstyle{{#1}\over{#2}}}}

\def\a{\alpha_0}

\def\del{\partial}

\def\acrit{\alpha_0^*}

\oneandahalfspace
\rightline{CTP TAMU--15/92}
\rightline{Preprint-KUL-TF-92/11}
\rightline{March 1992}

\vskip 2truecm
\centerline{\bf New Realisations of $W$ Algebras and $W$ Strings}
\vskip 1.5truecm
\centerline{H. Lu,$^*$ C.N. Pope,\footnote{$^*$}{\tenfoot {\sl Supported in
part by the U.S. Department of Energy, under
grant DE-FG05-91ER40633.}} S. Schrans\footnote{$^\diamond$}{\tenfoot {\sl
Onderzoeker I.I.K.W.;
On leave of absence from the Instituut voor Theoretische Fysica, \nl
\indent$\,$ K.U. Leuven, Belgium.}
}
and X.J.
Wang\footnote{}{\tenfoot }}
\vskip 1.5truecm
\centerline{\it Center
for Theoretical Physics,
Texas A\&M University,}
\centerline{\it College Station, TX 77843--4242, USA.}

\vskip 1.5truecm
\AB\singlespace
    We discuss new realisations of $W$ algebras in which the currents are
expressed in terms of two arbitrary commuting energy-momentum tensors
together with a set of free scalar fields.  This contrasts with the
previously-known realisations, which involve only one energy-momentum
tensor. Since realisations of non-linear algebras are not easy to come by,
the fact that this new class exists is of intrinsic interest. We use these
new realisations to build the corresponding $W$-string theories and show
that they are effectively described by two independent ordinary
Virasoro-like strings.
\AE\oneandahalfspace

\vskip 1.5truecm
\centerline{\tenfoot Available from hepth@xxx/9203024}

\np
\noindent
{\bf 1. Introduction}
\bigskip

     The quantisation of two-dimensional matter systems with local $W$
symmetries gives rise to new string theories, called $W$ strings [1,2,3,4,5,6].
Traditional Virasoro string theory acquires its physical interpretation when
the  Virasoro symmetry is realised by a set of free scalar fields, which then
describe the coordinates of the target spacetime.  Similarly, the physical
interpretation of a $W$ string depends upon the realisation of the $W$
symmetry.  For the bosonic $W\!A_n$ and $W\!D_n$ strings based on the $A_n$ and
$D_n$ classical simply-laced Lie algebras, the realisations that have been
found so far are given in terms of an arbitrary energy-momentum tensor $T^{\rm
eff}$, together with a set of free scalar fields $\varphi_i$ which are
``frozen'' by the physical-state conditions.  If $T^{\rm eff}$ is itself
realised in terms of free scalar fields $X^\mu$, then these scalars acquire the
interpretation of being target-spacetime coordinates.  The physical spectrum of
these $W$ strings thus closely resembles the spectrum of the ordinary Virasoro
string, but with a non-critical value of the central charge, and non-standard
values for the spin-2 intercept [5,6].

     In the first part of this paper we obtain new, inequivalent, realisations
for the $W\!A_n$ and $W\!D_n$ algebras with $n\ge 3$.  These realisations are
given in terms of {\it two} arbitrary energy-momentum tensors $T_+$ and $T_-$
(which commute with each other, and have the same central charge), together
with $(n-2)$ extra free scalar fields.  They are the first examples of
realisations that are not of the form described in the previous paragraph.
Since the $W$ algebras are non-linear, realisations are not easy to come by; in
particular, unlike the Virasoro algebra, one cannot obtain new realisations
simply by tensoring together old ones.  Thus these new realisations are of
intrinsic interest for their own sake.

     In the second part of this paper we use these realisations to construct
the corresponding $W$-string theories, and show that they are quite different
from the $W$-string theories that have previously been constructed.  The
physical spectrum is effectively described by two Virasoro-type strings living
in two independent Minkowski spacetimes.

\bigskip\bigskip
\noindent {\bf 2. Reduction}
\bigskip

     In order to explain how the new realisations of $W\!A_n$ and $W\!D_n$
arise, it is useful to begin by reviewing how the previously-known
realisations with only one arbitrary energy-momentum tensor are
constructed.\footnote{$^*$}{\tenfoot {\sl Recall that $W\!A_n$ is the $W_{n+1}$
algebra, generated by currents of spins $2,3,\ldots,n+1$ [7].  The $W\!D_n$
algebra is generated by currents of spins $2,4,6,\ldots,2n-2$ and $n$ [8].}}
The starting point is the Miura transformation, which provides a realisation
in terms of $n$ scalar fields [7,8].  For $W\!A_n$, $(n-1)$ of these scalars
appear in the generating currents only {\it via} the Miura transformation of
$W\!A_{n-1}$.  Thus the currents of $W\!A_n$ can be realised in terms of
those of $W\!A_{n-1}$, together with the remaining scalar field.  This
remarkable property is a consequence of the detailed form of the Miura
transformation.  It was first noticed for $W\!A_2=W_3$ in [9], and
generalised to all $W\!A_n$ in [2,5]. Applying this reduction recursively, one
can therefore realise the currents of $W\!A_n$ in terms of $(n-1)$ scalar
fields together with an arbitrary energy-momentum tensor $T^{\rm eff}$ which
generates $W\!A_1$, the Virasoro algebra.  This reduction procedure can be
summarised by the diagram
$$
W\!A_n\rightarrow W\!A_{n-1}\rightarrow\cdots\rightarrow W\!A_2 \rightarrow
W\!A_1=\hbox{Virasoro}.\eqno(2.1)
$$

     For $W\!D_n$, it was shown in [6] that a similar reduction is possible,
with the currents being expressed in terms of those of $W\!D_{n-1}$
together with an additional free scalar.  Since $D_2\cong A_1\times A_1$ is
not simple, one usually restricts the discussion of $W\!D_n$ algebras to the
cases $n\ge3$ [8].  Thus $W\!D_n$ can be realised in terms of the currents of
$W\!D_3$ together with $(n-3)$ free scalar fields.  Since $D_3\cong A_3$,
and thus $W\!D_3\cong W\!A_3$, we may then continue the reduction procedure
by following the route given in (2.1) to descend from $W\!A_3$ to $W\!A_1$.
This can be summarised by the diagram
$$
W\!D_n\rightarrow W\!D_{n-1}\rightarrow\cdots\rightarrow W\!D_3\cong W\!A_3
\rightarrow W\!A_2\rightarrow W\!A_1=\hbox{Virasoro}.\eqno(2.2)
$$
Thus $W\!D_n$ can be realised in terms of an arbitrary energy-momentum
tensor and $(n-1)$ additional free scalars.

     Although one does not usually consider the $W\!D_2$ algebra, the Miura
transformation for $W\!D_n$ allows one to follow the reduction scheme that
we have described above for one further step, to express the currents of
$W\!D_n$ in terms of those of $W\!D_2$ together with $(n-2)$ additional free
scalars.  Since $D_2\cong A_1\times A_1$, the algebra $W\!D_2$ is isomorphic
to the direct product of two independent Virasoro algebras.  Thus the
currents of $W\!D_n$ may be expressed in terms of $(n-2)$ free scalars
together with two arbitrary commuting energy-momentum tensors $T_+$ and
$T_-$.  The detailed form of the Miura transformation imposes the condition
that $T_+$ and $T_-$ should have the same central charge.  The reduction
procedure can be summarised by the diagram
$$
W\!D_n\rightarrow W\!D_{n-1}\rightarrow\cdots\rightarrow W\!D_3 \rightarrow
W\!D_2=\hbox{Virasoro}\times\hbox{Virasoro}.\eqno(2.3)
$$
Exploiting the isomorphism between $D_3$ and $A_3$ in the opposite direction
to that which we used in (2.2), we may also construct an alternative reduction
procedure for $W\!A_n$, given by
$$
W\!A_n\rightarrow W\!A_{n-1}\rightarrow\cdots\rightarrow W\!A_3\cong W\!D_3
\rightarrow W\!D_2=\hbox{Virasoro}\times\hbox{Virasoro}.\eqno(2.4)
$$
Thus the currents of $W\!A_n$ (with $n\ge3$) may be expressed in terms of
$(n-2)$ free scalars together with the two arbitrary energy-momentum tensors
$T_+$ and $T_-$.

     We shall now illustrate this reduction procedure in more detail. A
realisation of $W\!D_n$ in terms of $n$ free scalar fields
$(\varphi_1,\ldots,\varphi_n)$ has been given by Fateev and Lukyanov [8].
The spin-$n$ current is given by the Miura-type transformation
$$
f^{(n)}(z)=(\a\del-\del\varphi_n)\,(\a\del-\del\varphi_{n-1})\, \cdots\,
(\a\del-\del\varphi_2)\del\varphi_1\ , \eqno(2.5)
$$
where, as usual, normal ordering is assumed. The remaining currents
$W^{(n)}_{2k}$ of spin $2k$, with $k=1,2,\ldots,n-1$, can
then be read off from the poles of order $(2n-2k)$ in the operator-product
expansion $f^{(n)}(z)f^{(n)}(w)$ [8]. The reduction procedure
$W\!D_n\rightarrow
W\!D_{n-1}$ can now easily been understood by observing from (2.5) that we
may rewrite $f^{(n)}$ as
$$
f^{(n)}=\a\del f^{(n-1)}-\del\varphi_n f^{(n-1)}\ ,\eqno(2.6)
$$
where $f^{(n-1)}$ is the spin-$(n-1)$ current of $W\!D_{n-1}$. It then follows
that the $W^{(n)}_{2k}$ currents can, likewise, be written in terms of the
$W\!D_{n-1}$ currents together with the extra scalar field $\varphi_n$.
Applying this reduction procedure recursively one may then realise the $W\!D_n$
algebra in terms of the currents generating $W\!D_2$ together with the
$(n-2)$ free scalars $(\varphi_3, \ldots,\varphi_n)$.  As explained
previously, $W\!A_n$ can also be realised in terms of the currents
generating $W\!D_2$ together with $(n-2)$ free scalars, by following the
route given in (2.4).  Since the key step in these reduction schemes for
both $W\!A_n$ and $W\!D_n$ is the reduction from $W\!D_3$ (which is
isomorphic to $W\!A_3$) to $W\!D_2$, we shall focus for now on this
particular step.  This will then straightforwardly lead to new realisations
for $W\!A_n$ and $W\!D_n$.

    $W\!D_2$ is generated by a current $f^{(2)}$ of spin 2 given in (2.5) and
an  energy-momentum tensor $T^{(2)}=W^{(2)}_2$ with central charge $2(1+6\a
^2)$. Since $f^{(2)}$ turns out to be a spin-2 primary field with respect to
$T^{(2)}$, one can define two commuting spin-2 currents
$$
T_\pm= \ft12 \big( T^{(2)}\pm f^{(2)} \big)\eqno(2.7)
$$
which  both generate a Virasoro algebra with the same central charge
$$
c_\pm=1+6\a^2. \eqno(2.8)
$$

     It now follows from the reduction procedure (2.6)  that the currents of
$W\!A_3\cong W\!D_3$ can be written as
$$
\eqalign{
T=&\ T_+ + T_- -\ft12(\del\varphi_3)^2 + 2\a \del^2\varphi_3\ ,\cr
W=&\ \a\del( T_+- T_-)-\del \varphi_3(T_+ - T_-)\ ,\cr
U=&\  -(T_+ -T_-)^2 + 2 (\del \varphi_3)^2 (T_+ + T_-) -4 \a \del^2\varphi_3
(T_++T_-)
 - 2 \a \del \varphi_3 \del( T_+ + T_-)\cr
&\ - \a^2 \del^2(T_+ +  T_-)
+\ft12 (1+6\a^2) \del^3 \varphi_3 \del\varphi_3 -\ft23 \a (1+6\a^2)
\del^4 \varphi_3\ .\cr
}\eqno(2.9)
$$
Here $W=f^{(3)}$ is the spin-3 current given in (2.5), which is primary,
$T=W^{(3)}_2$ is the energy-momentum tensor and $U$ is the spin-4 current
which appears in the second order pole of the operator-product expansion
$W(z)W(w)$. (Note that $U$ is given in a different basis from the spin-4
current $W^{(3)}_4$ defined in [8]; they differ by a term proportional to
$\del ^2 T$.)  The term $(T_+-T_-)^2$ in $U$ is normal ordered with respect
to the modes of $T_+$ and $T_-$. One may check explicitly that these
currents generate the $W\!A_3\cong W\!D_3$ algebra with central charge
$$
c=3(1+20 \a^2).\eqno(2.10)
$$
The realisation (2.9) is remarkable in the sense that $T_+$ and $T_-$ may be
chosen to be {\it arbitrary} commuting energy-momentum tensors (with equal
central charges given by (2.8)). $\varphi_3$ is a free scalar which commutes
with $T_+$ and $T_-$. Since these energy-momentum tensors appear in the
combinations $T_++T_-$ as well as $T_+-T_-$, equation (2.9) thus gives a
realisation of $W\!A_3\cong W\!D_3$ in terms of two independent energy-momentum
tensors together with one free scalar field. As explained earlier, this
immediately yields realisations of $W\!A_n$ and $W\!D_n$ in terms of two
arbitrary (commuting) energy-momentum tensors $T_\pm$, with the same central
charge (2.8), together with $(n-2)$ free scalar fields.

     These new realisations (which are at the quantum level) imply in
particular the existence of new realisations of $W\!A_n$ and $W\!D_n$ at the
classical level, {\it i.e.}\ on the level of Poisson brackets. It has been
shown in [9] that Poisson-bracket realisations of $W\!A_2$ are related to
Jordan
algebras of cubic norm. This was generalized to a connection between
Poisson-bracket realisations of $W\!A_n$ and Jordan algebras of norm $\nu=n+1$
in [10], leading to a new Poisson-bracket realisation of $W\!A_n$, which is the
classical limit of the new (quantum) realisation presented in this paper.

\bigskip\bigskip
\noindent
{\bf 3. Applications to String Theory}
\bigskip

     We shall now use these realisations to construct new string theories with
a $W\!A_n$ or $W\!D_n$ symmetry.  We shall prove that these $W$ string theories
are related to Virasoro minimal models. In the case of realisations with
only {\it one} arbitrary energy-momentum tensor, it was shown in [5,6] that the
corresponding $W$-string theories are related to the $(h,h+1)$ unitary Virasoro
minimal models, where $h$ is the dual coxeter number of the underlying Lie
algebra $A_n$ or $D_n$. For the realisations with {\it two} arbitrary
energy-momentum tensors, the corresponding $W$-string theories turn out to be
related to two  copies of the $(h,h+1)$ unitary Virasoro minimal model.  We
shall treat the $W\!A_3\cong W\!D_3$ string in some detail; the generalisation
to $n \ge 4$ is then straightforward.

     The critical central charge for the $W\!A_3\cong W\!D_3$ string is
$c^*=246$. This implies that the critical value of the background-charge
parameter in the Miura transformation (2.5) is, using (2.10), given by
$(\acrit)^2=81/20$. The central charges of the energy-momentum tensors $T_\pm$
is then given by $c_\pm^*=26-\ft7{10}$. Here 26 is the critical central charge
of the ordinary Virasoro string and $\ft7{10}$ is the central charge of the
$(4,5)$ unitary Virasoro minimal model.  This connection is strengthened by
the fact that  the physical spectrum of the $W\!A_3\cong W\!D_3$ string is
related to this minimal model, as we now show.

     Physical states of the $W\!A_3\cong W\!D_3$ string are defined by the
conditions that they are annihilated by the positive Laurent modes of the
generating currents $T$, $W$ and $U$ and that they are eigenstates of the
zero Laurent modes with specific eigenvalues, {\it viz.}\ the intercepts
$\omega_2$, $\omega_3$ and $\omega_4$ respectively. In
principle these intercepts should be determined by a BRST analysis. A simpler
method to determine them is to assume the existence of a particular
``tachyonic'' physical state, the ``cosmological solution" [2]. It has been
proven for $W\!A_3\cong W\!D_3$,  and argued for $W\!A_n$ with $n\ge 4$, in [5]
that the string theory based on a realisation with only one arbitrary
energy-momentum tensor  is unitary if and only if this cosmological solution is
contained in its physical spectrum.  The intercepts can thus be read off from
the action of the currents  on this particular state, and are given by
$$
\omega_2=10, \qquad \omega_3=0 \qquad \hbox{and}\qquad \omega_4=-\ft{2607}5.
\eqno(3.2)
$$
Since these intercepts depend on the basis chosen to describe the algebra, some
of them differ from the ones presented in [5]. One can check, however, that
the intercept values (3.2) are consistent with the ones given in [5]. In
particular the spin-3 intercept in (3.2) is zero, in accordance with the fact
that a spin-3 primary current always has zero intercept for any $W\!A_n$
string [5].

      Having obtained the intercept values, we are now in a position to
construct the string theory based on this realisation. Let us consider the
energy-momentum tensors $T_{\pm}$ to be realised by $D_{\pm}$ free scalars
$\vec X_\pm$:
$$
T_\pm=-\ft12 (\del\vec X_\pm)^2+\vec Q_\pm\cdot\del^2\vec X_\pm\ ,\eqno(3.3)
$$
where the background-charge vectors $\vec Q_\pm$ are chosen so that (2.8) is
satisfied, {\it i.e.}\
$$
(\vec Q_\pm)^2=\ft1{12}\big( \ft{253}{10}-D_\pm\big)\ .\eqno(3.4)
$$
For reasons we shall explain presently, all physical operators are of the
form
$$
V=
R(\vec X_+,\vec X_-) e^{\vec\beta_+\cdot\vec X_+ +\vec\beta_-\cdot\vec X_-
+\beta_3\varphi_3} \ ,\eqno(3.5)
$$
where $R(\vec X_+,\vec X_-) $ is a differential polynomial in $\vec X_+$
and $\vec X_-$. Note in particular that physical states have no excitations
in the frozen $\varphi_3$ direction. Clearly any $R(\vec X_+,\vec X_-)$ can be
expressed as
$$
R(\vec X_+,\vec X_-) =\sum_{i=1}^k R_+^i(\vec X_+)\, R_-^i(\vec X_-)\ ,
\eqno(3.6)
$$
where $k$ is some integer. Each differential polynomial $R_\pm^i(\vec X_\pm)$
has level number $m_\pm^i$, defined by $(T_\pm)_0 R_\pm^i=m_\pm^i R_\pm^i$.
In order for (3.5) to satisfy the intercept conditions (3.2), one can show from
(2.9) that all $R_+^i$ must have the same level number $m_+$, and all $R_-^i$
must have the same level number $m_-$. The physical-state conditions
then imply that (3.5) must satisfy
$$
T_\pm(z)\, V(w)\sim{{\Delta_\pm V(w)}\over{(z-w)^2}}
+{{\del_\pm V(w)}\over{z-w}}
\ ,\eqno(3.7)
$$
where $\Delta_\pm$ are the effective intercepts for $T_\pm$, and $\del_+$
and $\del_-$ stand for derivatives $\del$ acting only on $X_+$ terms and
$X_-$ terms respectively. Note that although $V$ is not primary under $T_+$ or
$T_-$ separately, it is annihilated by all positive Laurent modes of
$T_\pm$. In fact, one can show that every physical operator $V$ can be
written as a sum of factorised products of the form
$$
V_+V_- e^{\beta_3\varphi_3}  \ , \eqno(3.8)
$$
where $V_+$ is built only from $\vec X_+$ and is primary under $T_+$ with
dimension $\Delta_+$ whilst $V_-$ is built only from $\vec X_-$ and is
primary under $T_-$ with dimension $\Delta_-$. Thus the theory has cloven
itself into two independent Virasoro-like string theories with
energy-momentum tensors $T_\pm$ having intercepts $\Delta_\pm$, and physical
operators $V_\pm$ .

    Introducing the shifted momentum $\gamma_3$ by
$\beta_3=\acrit (2+\gamma_3/9)$, the intercept conditions for the $W\!A_3\cong
W\!D_3$ string then give
$$
\eqalignno{
\big(T\big)_0:&\qquad 40(\Delta_+ +\Delta_-) -\gamma_3^2-76=0\ ,&(3.9a)\cr
\big(W\big)_0:&\qquad \gamma_3(\Delta_+ -\Delta_-)=0\ ,&(3.9b)\cr
\big(U\big)_0:&\qquad 200(\Delta_+-\Delta_-)^2 -20(\Delta_+
+\Delta_-)(\gamma_3^2-587) -11(2028+23\gamma_3^2)=0.&(3.9c)\cr}
$$
There are six solutions to these equations, namely
$$
\hbox{
\vbox{\tabskip=0pt \offinterlineskip
\def\tablerule{\noalign{\hrule}}
\halign to175pt{\strut#& \vrule#\tabskip=1em plus2em&
\hfil#\hfil& \vrule#& \hfil#\hfil& \vrule#&
\hfil#& \vrule#\tabskip=0pt\cr\tablerule
&&$\Delta_+$&&$\Delta_-$&&$\gamma_3$&\cr\tablerule
&&${\scriptstyle 9/10}$&&${\scriptstyle 1}$&&${\scriptstyle 0}$&\cr
\tablerule
&&${\scriptstyle 1}$&&${\scriptstyle 9/10}$&&${\scriptstyle 0}$&\cr
\tablerule
&&${\scriptstyle 77/80}$&&${\scriptstyle 77/80}$&&${\scriptstyle\pm 1}$&\cr
\tablerule  &&${\scriptstyle 1}$&&${\scriptstyle 1}$&&${\scriptstyle\pm 2}$&\cr
\tablerule \noalign{\smallskip}}}}\eqno(3.10)
$$
Thus we see that the $\Delta_\pm$ values can all be written in the form
$1-L_0^{\rm min}$, where $L_0^{\rm min}$ takes its values from the diagonal
entries $\Delta_{(r,r)}$ of the Kac table of the $(4,5)$ minimal model.

     It is now straightforward to demonstrate the unitarity of the theory
since it is effectively described by two independent Virasoro-like strings,
each with central charge $26-\ft7{10}$ and intercept values given in
(3.10).  Since these lie within the unitarity bounds for the $c=26-\ft7{10}$
string, the absence of ghosts is manifest.

     In the discussions so far we have only considered states that have no
excitations in the frozen direction $\varphi_3$. Following [5], we can show
that the momentum-conservation law in the $\varphi_3$ direction, which must
be satisfied in order to have non-vanishing two-point functions, requires
that the physical operators should come in conjugate pairs with equal and
opposite shifted momenta $\gamma_3$. For the physical operators we have
considered so far, this is indeed the case, as one can see in (3.10).
Explicit computations for level 1 states, combined with evidence found
previously for other $W$ strings [5,6], indicate that this conjugation
property is never satisfied for states with excitations in the frozen
direction. Thus such states have vanishing two-point functions and therefore
have zero norm. Consequently, the only physical states are indeed given by
sums of terms of the form (3.8).

\bigskip\bigskip
\noindent
{\bf 4. Conclusions}
\bigskip

     In this paper we have described new realisations for the $W\!A_n$ and
$W\!D_n$ algebras for $n\ge3$. The currents are realised in terms of two
arbitrary energy-momentum tensors together with $(n-2)$ free scalars. We
have discussed in detail the example of $W\!A_3\cong W\!D_3$, and given
the explicit form of the currents. We have used this realisation to
construct a new $W$ string. The essential new feature of this string theory
is that it is effectively reduced to two copies of Virasoro-like strings
with the same central charge $26-\ft7{10}$.
These strings are independent except that their intercepts are correlated,
as can be seen in (3.10).

     This effective theory is very similar to the theory that one would
obtain by starting from an action that was the sum of two independent
Virasoro-string actions. Thus it describes two ``non-communicating'' worlds,
each with its own spacetime.

      Although we have concentrated for simplicity on the $n=3$ case, the
generalisation to $W\!A_n$ and $W\!D_n$ for arbitrary $n$ is completely
straightforward, using the reduction procedures explained in section 2.
The intercepts for the two effective Virasoro-string theories can easily be
found from the action of the Weyl group of $A_n$ or $D_n$ on the
cosmological solution in the $n$-scalar realisation, using the same
techniques that were developed in [5,6].  These effective intercepts are
again related to the diagonal entries of the Kac tables for the $(h,h+1)$
unitary Virasoro minimal models, where $h$ is the dual Coxeter number of the
underlying Lie algebra.

\bigskip\bigskip\bigskip
\centerline{\bf ACKNOWLEDGMENTS}
\bigskip

Stany Schrans is much obliged to the Center for Theoretical Physics,
Texas A\&M University, for hospitality, and to the Belgian National Fund for
Scientific Research for a travel grant.

\bigskip\bigskip
\singlespace
\centerline{\bf REFERENCES}
\frenchspacing
\bigskip

\item{[1]}C.N.\ Pope, L.J.\ Romans and K.S.\ Stelle, {\sl Phys.\
Lett.}\ {\bf 268B} (1991) 167;\nl
{\sl Phys.\ Lett.}\ {\bf 269B} (1991) 287.

\item{[2]}S.R.\ Das, A.\ Dhar and S.K.\ Rama, {\sl Mod.\ Phys.\ Lett.}\
{\bf A6} (1991) 3055;\nl
``Physical states and scaling properties of $W$  gravities and $W$ strings,''
TIFR/TH/91-20.

\item{[3]}C.N.\ Pope, L.J.\ Romans, E.\ Sezgin and K.S.\ Stelle, {\sl Phys.\
Lett.}\  {\bf 274B} (1992) 298.

\item{[4]}S.K.\ Rama, {\sl Mod.\ Phys.\ Lett.}\ {\bf A6} (1991) 3531.

\item{[5]}H.\ Lu, C.N.\ Pope, S.\ Schrans and K.W.\ Xu, ``The Complete Spectrum
of the $W_N$ String,''  preprint CTP TAMU-5/92, KUL-TF-92/1.

\item{[6]}H.\ Lu, C.N.\ Pope, S.\ Schrans and X.J.\ Wang, ``On Sibling and
Exceptional $W$ Strings,''  preprint CTP TAMU-10/92, KUL-TF-92/8.

\item{[7]}V.A.\ Fateev and S.\ Lukyanov,  {\sl Int.\ J.\ Mod.\  Phys.}\ {\bf
A3} (1988) 507.

\item{[8]}S.L.\ Lukyanov and V.A.\ Fateev, {\sl Sov.\ Scient.\ Rev.}\ {\bf
A15} (1990) 1;\nl
{\sl Sov.\ J.\ Nucl.\ Phys.}\ {\bf 49} (1989) 925.

\item{[9]}L.J.\  Romans, {\sl Nucl.\  Phys.}\ {\bf B352} (1991) 829.

\item{[10]}C.M.\ Hull, {\sl Nucl.\  Phys.}\ {\bf B353} (1991) 707.

\bye